\documentclass[aps,prb,twocolumn,showpacs,superscriptaddress]{revtex4-1}
\usepackage{graphicx}
\usepackage{amssymb,amsmath}
\usepackage{color}
\begin{document}

\title{Annealing-induced magnetic moments detected by spin precession measurements in epitaxial graphene on SiC}
\author{Bastian Birkner}
\email{bastian.birkner@physik.uni-regensburg.de}
\affiliation{Institute of Experimental and Applied Physics, University of Regensburg, 93040 Regensburg, Germany}
\author{Daniel Pachniowski}
\affiliation{Institute of Experimental and Applied Physics, University of Regensburg, 93040 Regensburg, Germany}
\author{Andreas Sandner}
\affiliation{Institute of Experimental and Applied Physics, University of Regensburg, 93040 Regensburg, Germany}
\author{Markus Ostler}
\affiliation{Lehrstuhl f\"ur Technische Physik, University of Erlangen-N\"urnberg, 91058 Erlangen, Germany}
\author{Thomas Seyller}
\altaffiliation[Present address: ]{Technische Universit\"at Chemnitz, 09107 Chemnitz, Germany }
\affiliation{Lehrstuhl f\"ur Technische Physik, University of Erlangen-N\"urnberg, 91058 Erlangen, Germany}
\author{Jaroslav Fabian}
\affiliation{Institute of Theoretical Physics, University of Regensburg, 93040 Regensburg, Germany}
\author{Mariusz Ciorga}
\affiliation{Institute of Experimental and Applied Physics, University of Regensburg, 93040 Regensburg, Germany}
\author{Dieter Weiss}
\affiliation{Institute of Experimental and Applied Physics, University of Regensburg, 93040 Regensburg, Germany}
\author{Jonathan Eroms}
\affiliation{Institute of Experimental and Applied Physics, University of Regensburg, 93040 Regensburg, Germany}

\date{\today}
\begin{abstract}
We present results of non-local and three terminal (3T) spin precession measurements on spin injection devices fabricated on epitaxial graphene on SiC. The measurements were performed before and after an annealing step at 150~$^{\circ}$C for 15 minutes in vacuum. The values of spin relaxation length $L_s$ and spin relaxation time $\tau_s$ obtained after annealing are reduced by a factor 2 and 4, respectively, compared to those before annealing. An apparent discrepancy between spin diffusion constant $D_s$ and charge diffusion constant $D_c$ can be resolved by investigating the temperature dependence of the $g$-factor, which is consistent with a model for paramagnetic magnetic moments.
\end{abstract}
\maketitle
Apart from its prospects for electronic devices\cite{Number1,Number2} single layer graphene (SLG) is also a very promising candidate in the field of spintronics because it is expected that spin information can be passed in graphene over long distances\cite{Tombros} due to the weak spin-orbit coupling and low hyperfine interaction\cite{Huertas}. Up to now, however, the measured spin lifetimes in exfoliated SLG (0.5~ns at RT\cite{Han1}, $\approx1$~ns at 4~K\cite{Han2}) and also in bilayer graphene ($\approx$ 2~ns at RT\cite{aachen}) on SiO$_2$ are still one order of magnitude smaller than in conventional semiconductor heterostructures. Even if the mobility $\mu$ for graphene on SiO$_2$ is modified by e.g. ligand-bound nanoparticles\cite{HanMobility} (2700~cm$^2$/Vs~-~12\,000~cm$^2$/Vs) or by using high quality suspended graphene devices\cite{suspended} ($\mu>100\,000$~cm$^2$/Vs), measured spin lifetimes are below 2~ns.
Similar values of $\tau_s$, slightly over 2~ns, were also reported for graphene epitaxially grown on a semi-insulating silicon carbide (SiC) substrate\cite{First, Emtsev1} using a direct non-local measurement\cite{Maassen} while a huge $\tau_s$ was obtained by an indirect\cite{Fert} method. In Ref.~\onlinecite{Maassen} fitting the Hanle curves with a $g$-factor of 2 leads to a drastic difference between charge ($D_c$) and spin diffusion constant ($D_s$). Later the data were reinterpreted in a model employing a modified $g$-factor\cite{MaassenARXIV}. McCreary {\em et al.}\cite{Kawakamilocalmoments} studied the influence of artificially created paramagnetic moments on spin transport in graphene, and introduced an effective exchange field model leading to an enhanced $g$-factor. This variety of different results both at room and low temperature motivates further experiments on epitaxial graphene to understand the spin relaxation mechanism in order to control the spin information for future spintronic devices.

Here we also use epitaxial graphene grown on the Si face of SiC and present non-local and three terminal\cite{Dash} spin precession measurements. The latter probes the spin accumulation\cite{Fabian} directly underneath the injector electrode induced electrically by a spin polarized current. We compare the results before and after an annealing step and observe that our measurements after annealing can be well explained with an enhanced $g$-factor assuming that $D_c$ and $D_s$ are equal. As the temperature dependence of this increased $g$-factor shows a clear $1/T$ (paramagnetic) behavior, we believe that annealing creates local magnetic moments which influence the spin transport properties.
\begin{figure}
\centering
\includegraphics[width=7.5cm]{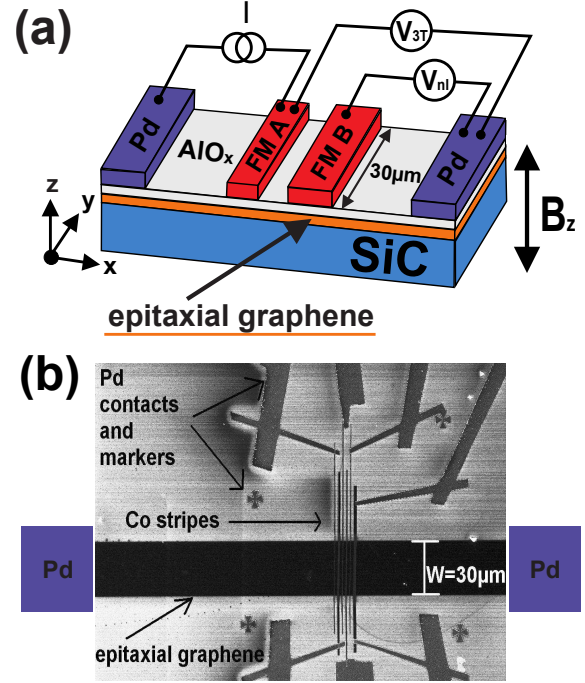}
\caption{(Color online) \textbf{(a)} Schematic drawing of the non-local and 3T measurement setup which is used for electrical spin injection and detection. The magnetic field B$_{z}$ is applied along the z axis. \textbf{(b)} SEM picture of the epitaxial graphene spin valve device.}\label{fig:sketch}
\end{figure}

Fig.~\ref{fig:sketch}{(a)} shows a sketch of the applied measurement methods, Fig.~\ref{fig:sketch}{(b)} a SEM picture of the used epitaxial graphene spin injection device. The graphene stripes having a width of $W=30~\mu$m and a length of about 750~$\mu$m are produced using a negative resist based electron beam lithography (EBL) step and oxygen plasma etching for 30~s (30~mTorr O$_2$, 50~W). Afterwards a thin tunneling barrier (AlO$_x$) with a thickness of about 1~nm was produced by depositing Al atoms over the entire cooled sample (180~K) in a UHV system ($p\approx10^{-9}$~mbar) and subsequent oxidation in the load lock in pure oxygen atmosphere ($p\approx3\times10^{-2}$~mbar) at RT for 30 minutes. This AlO$_x$ tunneling barrier with a contact resistance $R_c\geq2~k\Omega$ provides high spin injection efficiencies and reduces spin relaxation induced by the contacts\cite{Pop}. The ferromagnetic (FM) cobalt electrodes (Co 20~nm) with a width of 200~nm (contact A) and 500~nm (contact B) and the non-magnetic palladium contacts (Pd 80~nm) were each patterned using a positive PMMA resist based EBL step. The evaporation is done via electron gun (Co) and thermally (Pd) at a base pressure of about $5\times10^{-7}$~mbar followed by a standard lift-off technique. The distance $L$ between the edges of the FM stripes is 2~$\mu$m. Finally the sample is glued into a chip carrier and the measurements are done using a standard DC setup in a Cryogenics He-4 cryostat ($T=1.6\ldots300$~K) equipped with a vector magnet ($B_{x,y,z}=-1\ldots1$~T). The complete sample fabrication is done without applying a high temperature cleaning step.
\begin{figure}
  \centering
     \includegraphics[width=8.5cm]{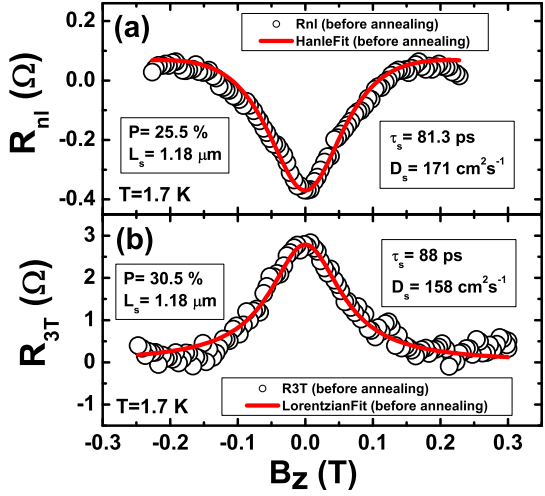}
    	\caption{(Color online) \textbf{(a)} Hanle spin precession measurement (background removed) with a DC current of +10~$\mu$A at 1.7~K and fit (continuous line) in non-local configuration. \textbf{(b)} 3T measurement (background removed) and Lorentzian fit (continuous line). Both measurements are done \textit{before} an annealing step.}\label{fig:before}
\end{figure}

In Fig.~\ref{fig:before} typical Hanle curves in the non-local and three terminal setup at 1.7~K are shown when using contact A as an injector. The FM stripes are magnetized in parallel configuration and the magnetic field $B_z$ is applied out-of plane which leads to dephasing of the spin signal. In our convention, $R_{nl}$ is negative for parallel magnetization.
The continuous curve for the non-local curve in Fig.~\ref{fig:before}{(a)} is the numerical fit with the solution of the following equation\cite{Ciorga}
\begin{equation}\label{nonlocal}
-R_\mathit{nl}=\frac{V_\mathit{nl}}{I}=\frac{P^2R_sL_s}{2W}\int_{0}^{\infty}\frac{cos(\omega_L t)}{\sqrt{4\pi D_s t}}e^{\frac{-(x_2-x_1)^2}{4D_s t}}e^{\frac{-t}{\tau_s}}dt
\end{equation}
where $P$ is assumed to be the same for both FM stripes and $x_1$, $x_2$ are the points of injection and detection, respectively. $P$ is the spin injection efficiency, $I$ the injection current, $\omega_L=g\mu_BB_z/\hbar$ the Larmor frequency with the Land\'{e} factor $g$, $R_s$ is the sheet resistance of graphene, $W$ the width of the graphene stripe, $D_s$ is the spin diffusion constant, finally $\tau_s$ and $L_s=\sqrt{D_s\tau_s}$ are the spin relaxation time and length, respectively. An influence of drift can be neglected due to the low bias current of 10~$\mu$A~\cite{Fabian,Kameno}.

The Hanle signal $R_\mathit{3T}$ in 3T configuration (Fig. \ref{fig:before}(b)) can be fitted with the following Lorentzian\cite{Dash} curve:
\begin{equation}\label{R3T}
R_\mathit{3T}=\frac{V_\mathit{3T}}{I}=\frac{P^2R_sL_s}{2W(1+(\omega_L\cdot\tau_s)^2)}
\end{equation}
We observe experimentally that $\tau_s$ from this fit coincides with $\tau_s$ obtained from the non-local measurement\cite{remark}. As the amplitude of the non-local signal is determined by the product of $P^2$ and $L_s$ these parameters are not independent in the fitting procedure. Therefore $L_s=1.18~\mu$m is estimated assuming an exponential decay of the spin signal given by $R_\mathit{3T}(B_z=0)$ at $ L=0~\mu$m and $R_\mathit{nl}(B_z=0)$ at $L=2~\mu$m. Now $P$ and $\tau_s$ are the free fitting parameters and $D_s$ can be calculated via $D_s=L^2_s/\tau_s$.

In Fig.~\ref{fig:before} one can see that for both non-local and 3T spin precession measurements the results for $P$, $\tau_s$ and $D_s$ are almost identical. This agreement indicates that these signals originate from an induced spin accumulation into graphene. Slight differences especially in the spin injection efficiency $P$ can be explained by a small anisotropic magneto resistance contribution of about 0.5~$\Omega$ of the FM stripes to $R_\mathit{3T}$ determined in reference measurements (not shown). This small deviation leads to an absolute error of $L_s$ of about $\Delta L_s=200$~nm. Fitting the non-local measurements, we obtain $\tau_s=81.3$~ps which is slightly smaller than in exfoliated SLG\cite{Tombros,Pop,Han}. The resulting spin diffusion constant $D_s=171$~cm$^2$/s is comparable to the charge diffusion constant $D_c=\frac{1}{2}l_pv_F=158$~cm$^2$/s extracted from a reference sample grown with identical parameters and also covered with AlO$_x$ produced with the same processing steps as for the tunneling barriers. This similarity shows that the value of $D_s$ extracted from the Hanle fit is reliable. $l_p=\frac{\hbar}{e}\sqrt{n\pi}\mu$ is the mean free path, $v_F=10^6$~m/s is the Fermi velocity in graphene, $n=5.9\times10^{12}$~cm$^{-2}$ is the charge carrier density and $\mu=1126$~cm$^2$/Vs is the mobility of the reference sample.

\begin{table}
\begin{tabular}{rclcrcrccc}
\hline\hline
\multicolumn{3}{c}{\textbf{Before annealing}} & & \multicolumn{6}{c}{\textbf{After annealing}}  \\
\cline{1-2}
\cline{2-3}
\cline{5-6}
\cline{6-7}
\cline{7-8}
\cline{8-9}
\cline{9-10}
$\tau_s$  &  $D_s$  &  $g_0$  &  \small\textbf{Injector}  &  $\tau_s$  &  $D_s$  &  $g_0$  \vline\vline\vline  &  $\tau_s$  &  $D_s$  &  $g_\mathit{eff}$  \\
\hline
81.3  \vline  &  171  \vline  &  2  &  A  &  95  \vline  &  37  \vline  &  2  \vline\vline\vline  &  22  \vline  &  160  \vline  &  8  \\
108  \vline  &  208  \vline  &  2  &  B  &  165  \vline  &  21  \vline  &  2  \vline\vline\vline  &  22  \vline  &  160  \vline  &  11  \\
\hline\hline
\end{tabular}
\caption{$D_s$~[cm$^2$/s], $\tau_s$~[ps] and $g$-factor before and after annealing for injector contact A and B at $T=1.7$~K. After annealing the measurements can also be fitted with an enhanced $g_\mathit{eff}>g_0$.}
\end{table}
In order to check if annealing influences the charge transport properties and/or the induced spin accumulation, a post annealing step is done at 150~$^{\circ}$C for 15 minutes in vacuum to avoid intercalation of hydrogen\cite{Speck,Riedl} via forming gas. Then we repeat the same spin precession measurements as before and interestingly we observe that $\tau_s$ increases whereas $D_s$ is decreased by almost a factor of 5 if we assume the same $g$-factor as before annealing ($g=g_0=2$). For the configuration with contact B as an injector we observe an even bigger decrease of $D_s$ by a factor of about 10 (Table I) which can be explained by an inhomogenity of $R_s$ after AlO$_x$ deposition also observed in the reference sample.
At this point we conclude that annealing affects the spin transport properties and we observe the same apparent reduction of $D_s$ as in Ref.~\onlinecite{Maassen}, where a high temperature annealing step was included in the sample preparation procedure. $L_s=594$~nm is reduced by a factor of 2 after annealing and is again extracted from the exponential decay of the spin signal at the injection point ($R_\mathit{3T}(0)$) and at a distance $L=2$~$\mu$m ($R_{nl}(0)$). The fact that both $L_s$ and the 3T amplitude at zero magnetic field decrease by almost the same factor indicates that the applied post annealing step also affects the induced spin accumulation underneath the injector electrode.
As the spin transport sample did not allow us to determine the mobility and charge carrier density independently we also annealed the reference sample (covered with AlO$_x$) under the same conditions as the spin transport sample. From low field Hall measurements at 1.7~K we get an enhanced charge carrier density $n=8.4\times10^{12}$~cm$^{-2}$ after annealing whereas the mobility just slightly increases to $\mu=1237$~cm$^2$/Vs. We conclude now that a change in $R_s$ is mainly caused by a change in the charge carrier density. Following the results of the reference sample we ascribe the small sheet resistance increase from $R_s=1.5$~k$\Omega$ before and $R_s=1.7$~k$\Omega$ after annealing of the spin transport sample to a minute reduction in doping. The charge diffusion constant $D_c$ (and also $D_s$) is therefore slightly decreased from 171~cm$^2$/s to 160~cm$^2$/s ($D_c\propto\sqrt{n}$) for the spin transport sample\cite{Supplemental}. In conclusion, the minor decrease in $D_c$ due to the annealing step cannot explain the strong reduction of $D_s$ extracted from the Hanle fits. That means we have the following situation: $D^\mathit{before}_c\approx D^\mathit{after}_c\gg D^\mathit{after}_s$.

In an attempt to understand the discrepancy of $D_c$ and $D_s$ Maassen {\em et al.}\cite{Maassen} first considered localized states in the electrically inert buffer layer\cite{Emtsev2} (BL) which could provide hopping sites for the spins being able to change the spin transport properties but not the charge transport properties. The difference in $D_s$ and $D_c$ was also recently discussed by McCreary {\em et al.}\cite{Kawakamilocalmoments}. They assume a formation of local magnetic moments by Ar sputtering or from hydrogen adatoms on exfoliated graphene samples which provide an enhanced magnetic field for the diffusing spins, which can be modeled by an effective $g$-factor.
Maassen {\em et al.}\cite{MaassenARXIV} reinterpreted their experiments\cite{Maassen} using a model of localized states, where the effective Larmor frequency is increased in the limit of strong coupling, which again can be expressed by an enhanced $g$-factor and then allows to set $D_c=D_s$.
\begin{figure}
  \centering
    \includegraphics[width=8.5cm]{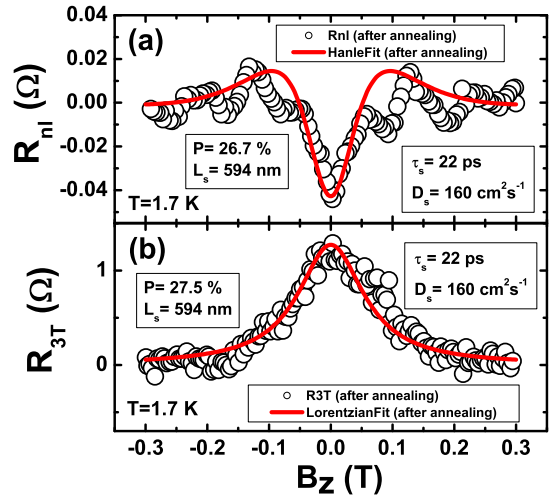}
   \caption{(Color online) \textbf{(a)} Hanle spin precession measurement (background removed) at 1.7~K and fit (continuous line) in non-local configuration. \textbf{(b)} 3T measurement (background removed) and Lorentzian fit (continuous line). Both measurements are done \textit{after} an annealing step. The fittings are done treating the $g$-factor as a free parameter.} \label{fig:after}
	\end{figure}

For this reason we also fit our non-local and 3T data after annealing, treating the $g$-factor in the Larmor frequency (Eqs. (\ref{nonlocal}) and (\ref{R3T})) as a free parameter, and assuming $D^\mathit{after}_c=D^\mathit{after}_s=160~$cm$^2$/s. Fig. \ref{fig:after} shows that our data can also be well fitted with an enhanced $g$-factor of 8 in both the non-local and in the 3T setup. The oscillations observed in the Hanle curve (Fig. \ref{fig:after}{(a)}) at higher magnetic fields are phase coherent contributions and vanish at higher temperatures.
If we summarize our experimental findings so far (Table I) we can conclude that our measurements after annealing can be explained either by $D^\mathit{after}_c\neq D^\mathit{after}_s$ or by an effective Land\'{e} factor $g_\mathit{eff}>2$ as both models reproduce the data equally well since Eqs. (\ref{nonlocal}) and (\ref{R3T}) contain the $g$-factor implicitly in the Larmor frequency $\omega_L$ and are invariant under a rescaling of $g$, $\tau_s$ and $D_s$~\cite{MaassenARXIV}.

\begin{figure}
  \centering
 \includegraphics[width=8.5cm]{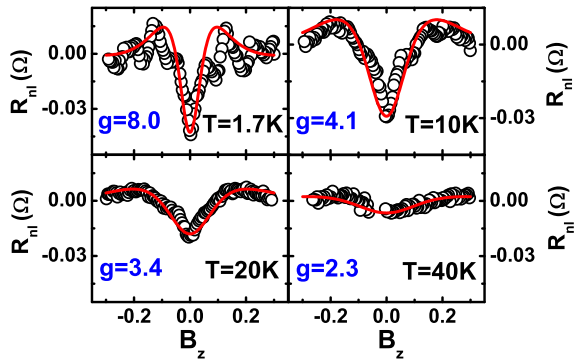}
    		\caption{(Color online) Temperature dependent non-local Hanle measurements (background removed) with $g$-factor as further fitting parameter.}\label{fig:T}
	\end{figure}

To determine which model (hopping or magnetic moments) is appropriate in our situation we study the temperature dependence of spin transport. We observe that the enhanced effective $g$-factor, as well as the amplitude of the spin signal, decrease with increasing temperature. From the $T$ dependence of the reference sample and the $T$ dependence of $R_s$ of the spin transport sample after annealing we conclude that $D_c$ is weakly influenced by the temperature. This was also included in the Hanle fits (Fig. \ref{fig:T}).
	
\begin{figure}[h!]
  \centering
 \includegraphics[width=8.4cm]{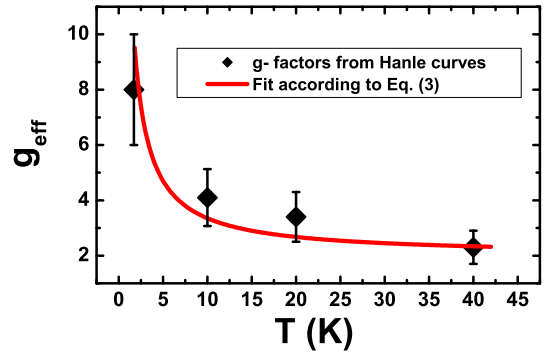}
    		\caption{(Color online) Temperature dependence of the $g$-factor. The continuous curve is the fit according to Eq. (\ref{gfactor}).}\label{fig:TAbfall}
	\end{figure}

If $g_\mathit{eff}$ originates from magnetic moments then its temperature dependence can be described by the following equation\cite{Kawakamilocalmoments}:
\begin{equation}\label{gfactor}
g_\mathit{eff}(T)=g_0+\frac{g_0\eta_M A_\mathit{ex}}{k_BT}
\end{equation}
This is the low field approximation of the Brillouin function of a spin $1/2$ paramagnetic material. $A_\mathit{ex}$ is the strength of the exchange coupling, $\eta_M$ represents the filling density of the magnetic moments, $g_0=2$ is the $g$-factor for free electrons and $k_B$ is the Boltzmann constant.		
As it is seen in Fig. \ref{fig:TAbfall} the measured temperature dependence of $g_\mathit{eff}$ is well described by Eq. (\ref{gfactor}). This temperature dependence is compatible with the effective exchange field model proposed by McCreary {\em et al.}\cite{Kawakamilocalmoments} which describes the enhancement of the magnetic field felt by the diffusing spins due to localized paramagnetic moments. Maassen {\em et al.}\cite{MaassenARXIV}, on the other hand, interpret their data by hopping of the diffusing spins into localized states which leads to an apparent enhancement of the $g$-factor in the Hanle fit. In their work the increase of $g$ is most pronounced at room temperature, whereas in our case the maximum $g_\mathit{eff}$ is obtained at low temperature. We therefore believe that post annealing creates an amount of randomly positioned magnetic moments resulting in an increased effective magnetic field $B_\mathit{eff}=B_z+B_\mathit{ex}$ composed of the applied out-of plane magnetic field $B_z$ and of an exchange field $B_\mathit{ex}$ coming from the induced magnetic moments\cite{Supplemental2}. This enhanced magnetic field can be modeled by an effective $g$-factor in the Larmor frequency $\omega_L$. As we nearly get the same Land\'{e} factor from the non-local Hanle and Lorentzian fit we can not decide if the magnetic moments are formed in the graphene/buffer layer transition or at the AlO$_x$/graphene interface. The difference between our experiment and the work of Maassen {\em et al.}\cite{Maassen} may be due to different annealing conditions.

As to the origin of the magnetic moments we assume that defects or vacancies\cite{Yazyev} which are already present in our epitaxial graphene are modified via the annealing step at 150~$^{\circ}$C. One example could be step edges, which are known to occur frequently in epitaxial graphene on SiC\cite{Emtsev1}. This is supported by weak localization measurements on the reference sample, which yield a very short intervalley scattering length of $L_i\approx40$~nm and also by THz photocurrent experiments on the reference sample where photocurrents were detected in the bulk of the sample\cite{Ganichev1,Ganichev2} at normal incidence, which can only be explained by a lowering of the symmetry\cite{Ganichev3}. Annealing then only change the termination of the step edges, which influences their magnetic behavior.
	
In conclusion, an electrically induced spin imbalance from ferromagnetic Co stripes can be analyzed via spin precession measurements in both non-local and three terminal configuration. By introducing a post annealing step, we observe that the spin relaxation length as well as the non-local and 3T Hanle amplitude decrease. Fitting of the non-local and 3T data after annealing shows an increase of the $g$-factor if spin and charge diffusion constants are assumed to be the same. The origin of the $g$-factor enhancement are local magnetic moments formed by annealing. The reduced spin lifetime and length support this assumption because local magnetic moments act as an additional spin scattering source. Finally, the temperature dependence shows a clear evidence that paramagnetic moments are created as the effective $g$-factor scales with $1/T$ with increasing temperature.

Support from the DFG within SFB 689 ``Spin phenomena in reduced dimension'' and SPP 1459 ``Graphene'' is gratefully acknowledged. We would like to thank F. Fromm, T. Maassen, S. Ganichev, and R. Kawakami for helpful discussions and
T. Korn, F. Yaghobian, and C. Sch\"{u}ller for supporting Raman measurements.


\newpage
\onecolumngrid
{\Large \bf \hfill Supplemental Material\hfill}
\vspace{2cm}


We report here on magnetotransport measurements of the reference sample before and after an annealing step and discuss the observability of a dip in the non-local in plane measurements.\vspace{1cm}
\twocolumngrid

\begin{figure}
	\centering
		\includegraphics[width=0.5\textwidth]{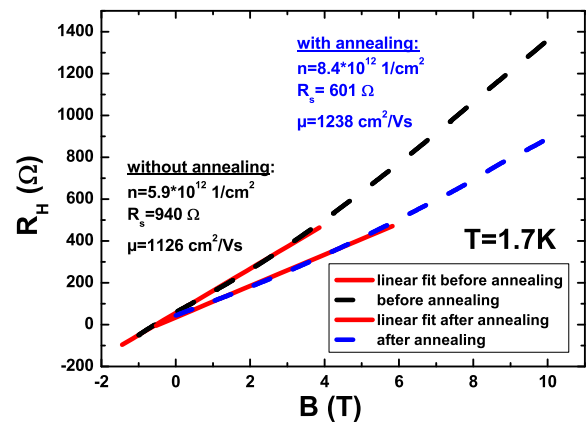}
	\caption{Magnetotransport measurement of the reference sample before and after an annealing step.}
	\label{fig:SHall}
\end{figure}

The strong modification of the Hanle curves before and after annealing can be explained by either an enormous reduction of the diffusion constant after annealing or by the creation of localized moments, resulting in a strongly enhanced $g$-factor of the electrons. To check that annealing barely changes the transport properties, we characterized the reference sample with magnetotransport measurements before and after an annealing step at 150~$^{\circ}$C for 15 minutes in vacuum. From these measurements we can extract the charge carrier density $n_s$ and the mobility $\mu$ in order to check the influence of annealing on these parameters. To this end we applied a perpendicular magnetic field and measured the Hall resistance R$_H$ at 1.7~K. We also determined the sheet resistance R$_s$ of graphene by applying the van der Pauw method. Finally, using the Drude formula, we calculated the carrier mobility $\mu=(R_sn_se)^{-1}$. Both carrier density and mobility before and after annealing are given as insets in Fig.~\ref{fig:SHall}.

From this study we conclude that annealing can indeed change the charge carrier density and the sheet resistance slightly but the mobility is almost unaffected. That means that a change in the sheet resistance results mainly from a change in the charge carrier density. Taking the results from the reference sample we draw the following conclusion for the sample in which spin transport was studied. In this the sheet resistance increased slightly upon annealing which can be now ascribed to a minute reduction in the charge carrier density. Therefore the charge diffusion constant for our spin transport sample is only slightly decreased as $D_c\propto\sqrt{n_s}$ and can not explain the strong reduction of the spin diffusion constant extracted from the Hanle data when using the $g$-factor for a free electron.

\begin{figure}
	\centering
		\includegraphics[width=0.5\textwidth]{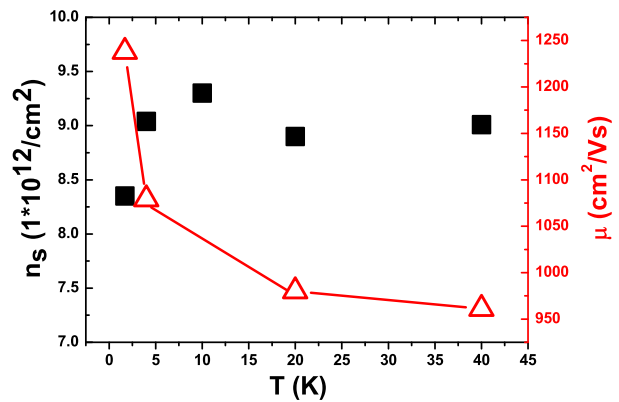}
	\caption{Charge carrier density $n_s$ and mobility $\mu$ versus temperature $T$ of the reference sample.}
	\label{fig:S2}
\end{figure}

We also performed Hall measurements for a number of temperatures after the annealing step, and determined the mobility $\mu$ at $T=1.7\ldots40$~K. From those $T$-dependent measurements we conclude that the carrier density $n_s$ is barely $T$-dependent but the mobility increases as the temperature is lowered (Fig.~\ref{fig:S2}). This information, together with the $T$-dependence of the sheet resistance of the spin transport sample allows us to model the $T$-dependence of the diffusion constant, which we need to determine the $T$-dependence of the $g$-factor.

Furthermore we discuss the dip feature around $B=0$ in the non-local spin valve measurements which was experimentally observed in Ref.~\onlinecite{Kawakamilocalmoments}.
Kawakami {\em et al.} observe a pronounced dip in the non-local resistance when they sweep the magnetic field in the direction along the ferromagnetic electrodes. This feature can be explained assuming the effective exchange-field model where the spin relaxation time $T^\mathit{total}_1$ is modeled in the following way:
\begin{equation}
\frac{1}{T^\mathit{total}_1}=\frac{1}{\tau^{so}}+\frac{1}{\tau^{ex}_1}
\label{Eq:T1total}
\end{equation}
where
\begin{equation}
\frac{1}{\tau^{ex}_1} = \frac{\frac{(\Delta B)^2}{\tau_c}\left(\frac{g_e}{g_e^*}\right)^2}{(B_\mathit{app,y})^2+\left(\frac{\hbar}{g_e^*\mu_B \tau_c}\right)^2}
\end{equation}

Here, $g_e^*$ is the effective $g$-factor, $g_e=2$ is the bare electron $g$-factor, $\Delta B$ and $\tau_c$ are the fluctuation amplitude and correlation time of the fluctuating field, felt by the moving electron spins, and $B_\mathit{app,y}$ is the external field, applied in the direction of the ferromagnetic electrodes.
In the case of Kawakami's experiment, the $B$-independent spin lifetime $\tau^{so}$ obtained from non-local Hanle measurements is quite long so an ensemble of fluctuating spins with $\Delta B=6.78$~mT and $\tau_c=192$~ps leads to a sizeable reduction of $T^\mathit{total}_1$.

\begin{figure}
	\centering
		\includegraphics[width=0.5\textwidth]{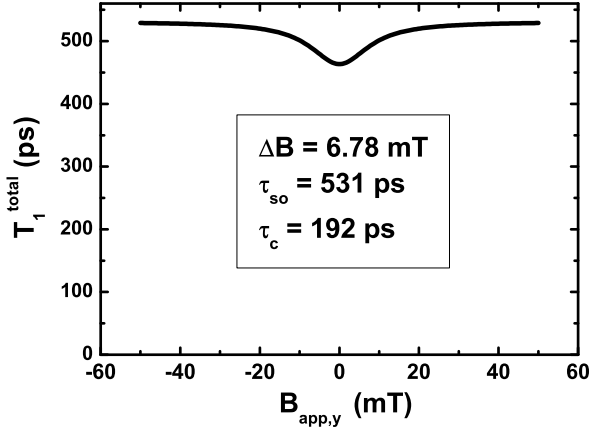}
	\caption{$T^\mathit{total}_1$ from Kawakami's group.}
	\label{fig:S3}
\end{figure}

\begin{figure}[h!]
	\centering
		\includegraphics[width=0.5\textwidth]{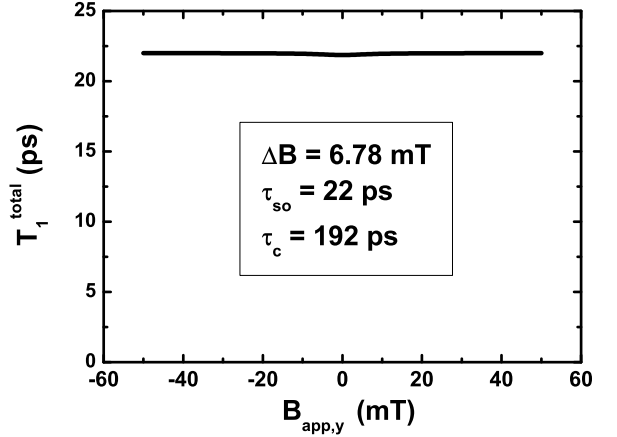}
	\caption{$T^\mathit{total}_1$ with $\tau^{so}$ from our data.}
	\label{fig:S4}
	\end{figure}

Since the spin-flip length is $L_s = \sqrt{D_s\,T^\mathit{total}_1}$ and the non-local signal in the tunneling regime\cite{Pop,Han2010} depends on $L_s$ according to
\begin{equation}\label{Rnl}
R_\mathit{nl}=\frac{P^2R_sL_s}{2W}exp(-L/L_s) ,
\end{equation}
a variation of $T^\mathit{total}_1$ directly modifies the observed non-local signal. In our case, the spin lifetime before and after annealing is 80 ps and 22 ps, respectively. Annealing creates localized moments, reducing the $B$-independent spin lifetime (modeled by $\tau^{so}$ in Eq. (\ref{Eq:T1total})) due to spin-flip scattering with the localized moments, while the additional, $B$-dependent reduction resulting from the effective exchange field model is barely visible. This is shown in Fig.~\ref{fig:S3} and Fig.~\ref{fig:S4}.
	
\begin{figure}
	\centering
		\includegraphics[width=0.5\textwidth]{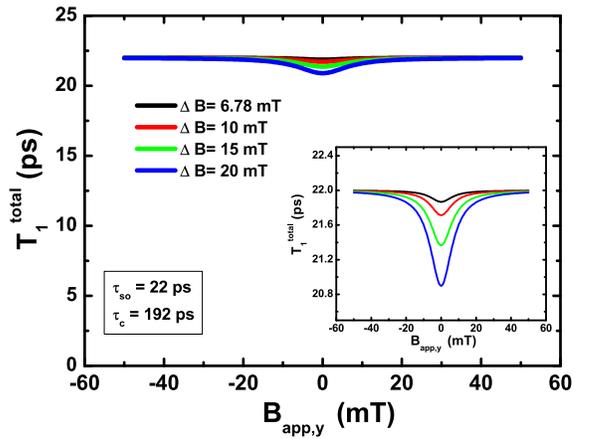}
	\caption{$T^\mathit{total}_1$ for different $\Delta B$.}
	\label{fig:S5}
\end{figure}

\begin{figure}
	\centering
		\includegraphics[width=0.5\textwidth]{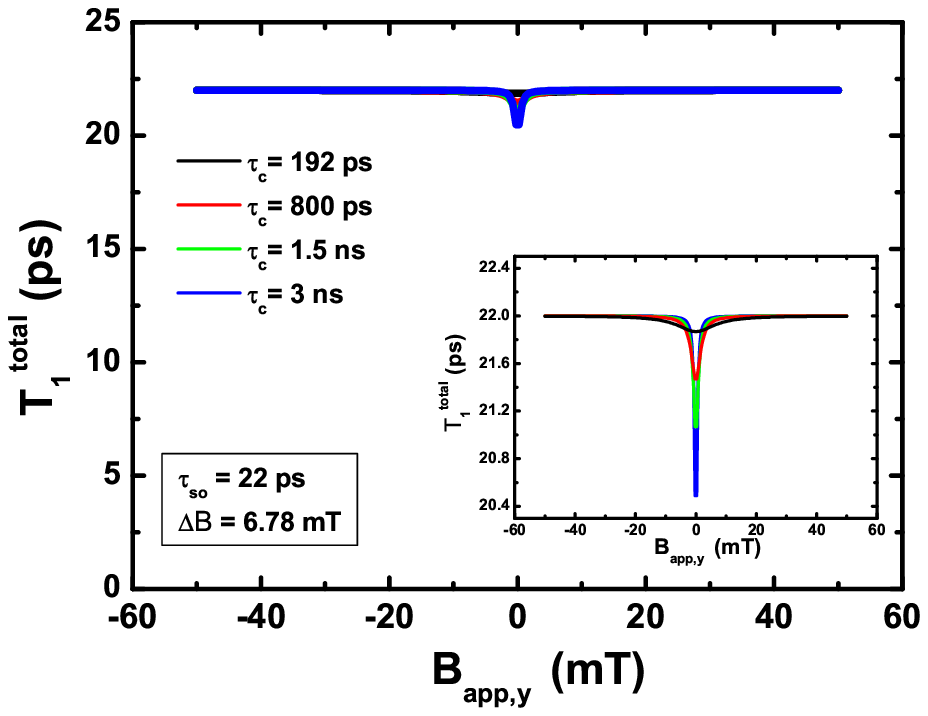}
	\caption{$T^\mathit{total}_1$ for different $\tau_c$.}
	\label{fig:S6}
\end{figure}

In Fig.~\ref{fig:S3} we reproduce the expected $T^\mathit{total}_1$-time for the experiment of the Kawakami group, showing a pronounced dip in $T^\mathit{total}_1$ around $B=0$.
If we replace $\tau^{so}$ with the experimental value for our sample after annealing, we find that for the same $\Delta B=6.78$~mT and $\tau_c=192$~ps as in Kawakami's experiment the dip feature is now almost invisible (Fig.~\ref{fig:S4}).
\newline This also holds if we increase $\Delta B$ to 20~mT and $\tau_c$ to 3~ns (Fig.~\ref{fig:S5} and Fig.~\ref{fig:S6}). The corresponding tiny change in $R_{nl}$ of at most a few m$\Omega$ would be  unobservable, given the noise level of about 10 m$\Omega$. Indeed, experimentally we did not observe such a feature.


\begin{thebibliography}{99}
\bibitem{Number1}
A. H. Castro Neto, F. Guinea, N.M.R. Peres, K.S. Novoselov, and A. K. Geim, Rev. Mod. Phys. \textbf{81}, 109 (2009).
\bibitem{Number2}
K. S. Novoselov, A. K. Geim, S. V. Morozov, D. Jiang, M. I. Katsnelson, I. V. Grigorieva, S. V. Dubonos, and A. A. Firsov, Nature (London) \textbf{438}, 197 (2005).
\bibitem{Tombros}
N. Tombros, C. J\'{o}zsa, M. Popinciuc, H. T. Jonkman, and Bart J. van Wees, Nature (London) \textbf{448}, 571 (2007).
\bibitem{Huertas}
D. Huertas-Hernando, F. Guinea, and A. Brataas, Phys. Rev. B \textbf{74}, 155426 (2006).
\bibitem{Han1}
W. Han, K. Pi, K. M. McCreary, Y. Li, J. J. I. Wong, A. G. Swartz, and R. K. Kawakami, Phys. Rev. Lett. \textbf{105}, 167202 (2010).
\bibitem{Han2}
W. Han, and R. K. Kawakami, Phys. Rev. Lett. \textbf{107}, 047207 (2011).
\bibitem{aachen}
T.-Y. Yang, J. Balakrishnan, F. Volmer, A. Avsar, M. Jaiswal, J. Samm, S.R. Ali, A. Pachoud, M. Zeng, M. Popinciuc, G. G\"{u}ntherodt, B. Beschoten, and B. \"{O}zyilmaz, Phys. Rev. Lett. \textbf{107}, 047206 (2011).
\bibitem{HanMobility}
W. Han, J.-R. Chen, D. Wang, K. M. McCreary, H. Wen, A. G. Swartz, J. Shi, and R. Kawakami, Nano Lett. \textbf{12}, 3443 (2012).
\bibitem{suspended}
M. H. D. Guimar\~{a}es, A. Veligura, P. J. Zomer, T. Maassen, I. J. Vera-Marun, N. Tombros, and B. J. van Wees, Nano Lett. \textbf{12}, 3512 (2012).
\bibitem{First}
P. N. First, P. N., W. A. de Heer, T. Seyller, C. Berger, J. A. Stroscio, J.-S. Moon, MRS Bull. \textbf{35}, 296 (2010).
\bibitem{Emtsev1}
K. V. Emtsev, A. Bostwick, K. Horn, J. Jobst, G. L. Kellogg, L. Ley, J. L. McChesney, T. Ohta, S. A. Reshanov, J. R\"{o}hrl, E. Rotenberg, A. K. Schmid, D. Waldmann, H. B. Weber, and T. Seyller, Nat. Mat. \textbf{8}, 203 (2009).
\bibitem{Maassen}\label{Maassen}
T. Maassen, J. J. van den Berg, N. IJbema, F. Fromm, T. Seyller, R. Yakimova, and B. J. van Wees, Nano Lett. \textbf{12}, 1498 (2012).
\bibitem{Fert}
B. Dlubak, M.-B Martin, C. Deranlot, B. Servet, S. Xavier, R. Mattana, M. Sprinkle, C. Berger, W. A. De Heer,	F. Petroff,	A. Anane,	P. Seneor, and A. Fert, Nature Physics \textbf{8}, 557 (2012).
\bibitem{MaassenARXIV}
 T. Maassen, J. J. van den Berg, E. H. Huisman, H. Dijkstra, F. Fromm, T. Seyller and B. J. van Wees, arXiv:1208.3129v1 (2012).
 \bibitem{Kawakamilocalmoments}
K. M. McCreary, A. G. Swartz, W. Han, J. Fabian and R. K. Kawakami, Phys. Rev. Lett. \textbf{109}, 186604 (2012).
\bibitem{Dash}
S. P. Dash, S. Sharma, R. S. Patel, M. P. de Jong, and R. Jansen, Nature \textbf{462}, 491 (2009).
\bibitem{Fabian}
J. Fabian, A. Matos-Abiague, C. Ertler, P. Stano, and I. Zutic, Acta Phys. Slov. \textbf{57}, 565 (2007).
\bibitem{Pop}
M. Popinciuc, C. J\'{o}zsa, P. J. Zomer, N. Tombros, A. Veligura, H. T. Jonkman, and B. J. van Wees, Phys. Rev. B \textbf{80}, 214427 (2009).
\bibitem{Ciorga}
M. Ciorga, A. Einwanger, U. Wurstbauer, D. Schuh, W. Wegscheider, and D. Weiss, Phys. Rev. B \textbf{79}, 165321 (2009).
\bibitem{Kameno}
M. Kameno, Y. Ando, E. Shikoh, T. Shinjo, T. Sasaki, T. Oikawa, Y. Suzuki, T. Suzuki, and M. Shiraishi, Appl. Phys. Lett. \textbf{101}, 122413 (2012).
\bibitem{remark}
Strictly speaking, the Lorentzian is only obtained for contacts much wider than the spin relaxation length. In our case, this may lead to an underestimated $\tau_s$ by a factor of 2.
\bibitem{Han}
W. Han, and R. K. Kawakami, Phys. Rev. Lett. \textbf{107}, 047207 (2011).
\bibitem{Riedl}
C. Riedl, C. Coletti, T. Iwasaki, A. A. Zakharov, and U. Starke, Phys. Rev. Lett. \textbf{103}, 246804 (2009).
\bibitem{Speck}
F. Speck, J. Jobst, F. Fromm, M. Ostler, D. Waldmann, M. Hundhausen, H. B. Weber, T. Seyller, Appl. Phys. Lett. \textbf{99}, 122106 (2011).
\bibitem{Supplemental}
See supplemental material for control measurements and more details on the effective exchange field model.
\bibitem{Emtsev2}
K. V. Emtsev, F. Speck, T. Seyller, L. Ley and J. D. Riley, Phys. Rev. B \textbf{77}, 155303 (2008).
\bibitem{Supplemental2}
For further discussion regarding the dip feature around $B=0$ when $B$ is applied along the ferromagnetic stripes see Ref. \onlinecite{Supplemental}.
\bibitem{Yazyev}
O. V. Yazyev, and L. Helm,  Phys. Rev. B \textbf{75}, 125408 (2007).
\bibitem{Ganichev1}
J. Karch, C. Drexler, P. Olbrich, M. Fehrenbacher, M. Hirmer, M. M. Glazov, S. A. Tarasenko, E. L. Ivchenko, B. Birkner, J. Eroms, D. Weiss, R. Yakimova, S. Lara-Avila, S. Kubatkin, M. Ostler, T. Seyller, and S. D. Ganichev, Phys. Rev. Lett. \textbf{107}, 276601 (2011).
\bibitem{Ganichev2}
J. Karch, P. Olbrich, M. Schmalzbauer, C. Zoth, C. Brinsteiner, M. Fehrenbacher, U. Wurstbauer, M. M. Glazov, S. A. Tarasenko, E. L. Ivchenko, D. Weiss, J. Eroms, R. Yakimova, S. Lara-Avila, S. Kubatkin, and S. D. Ganichev, Phys. Rev. Lett. \textbf{105}, 227402 (2010).
\bibitem{Ganichev3}
S. D. Ganichev, private communication.
\bibitem{Han2010}
W. Han, K. Pi, K. M. McCreary, Y. Li, J. J. I. Wong, A. G. Swartz, and R. K. Kawakami, Phys. Rev. Lett. \textbf{105}, 167202 (2010).
\end{thebibliography}
\end{document}